\documentclass[fleqn,12pt]{wlscirep}
\usepackage[utf8]{inputenc}
\usepackage[T1]{fontenc}
\usepackage{lineno}
\title{Can gender inequality be created without inter-group discrimination?}
\author[1,*]{Sylvie Huet}
\author[2]{Floriana Gargiulo}
\author[3]{Felicia Pratto}
 
\affil[1]{Universit\'e Clermont Auvergne, INRAE, LISC, Aubi\`ere, 63170, France, and LAPSCO, Clermont-Ferrand, 63000, France}
\affil[2]{CNRS, Gemass, Paris, 75017, France}
\affil[3]{University of Connecticut, Dept of Psychological Sciences, 406 Babbidge Road, Unit 1020
Storrs, CT 06269, USA}
 
\affil[*]{sylvie.huet@inrae.fr}
 
\begin{abstract}
Understanding human societies requires knowing how they develop gender hierarchies which are ubiquitous. We test whether a simple agent-based dynamic process could create gender inequality. Relying on evidence of gendered status concerns, self-construals, and cognitive habits, our model included a gender difference in how responsive male-like and female-like agents are to others’ opinions about the level of esteem for someone. We simulate a population who interact in pairs of randomly selected agents to influence each other about their esteem judgments of self and others. Half the agents are more influenced by their relative status rank during the interaction than the others. Without prejudice, stereotypes, segregation, or categorization, our model produces inter-group inequality of self-esteem and status that is stable, consensual, and exhibits characteristics of glass ceiling effects. Outcomes are not affected by relative group size. We discuss implications for group orientation to dominance and individuals’ motivations to exchange.
\end{abstract}
\begin{document}
 
\flushbottom
\maketitle
% * <john.hammersley@gmail.com> 2015-02-09T12:07:31.197Z:
%
%  Click he title above to edit the author information and abstract
%
\thispagestyle{empty}
 
%\noindent Please note: Abbreviations should be introduced at the first mention in the main text – no abbreviations lists. Suggested structure of main text (not enforced) is provided below.
% \linenumbers 
\section*{Introduction}

Most large human societies are structured as group-based dominance hierarchies that include gender inequality, such that men have more status and power than women; these hierarchies sometimes also include another form of inequality such as those based on ethnicity, citizenship status, race, religion or sect, social class, economic role \cite{SidaniusPratto1999}. A fundamental question is how these group hierarchies develop.

Numerous theoretical processes have been proposed to explain the origins of gender inequality, and they are not mutually exclusive. For example, sexual strategies theory suggests that men will strive to obtain greater status than other men, and women will select mates with higher status \cite{Buss1993}. Cultural materialism theory argues that when an ecology makes it easier for men than for women and children to produce food, men will gain more status than women, compared to ecologies in which there is no gender advantage in food procurement \cite{Harris1993}. Whenever men can monopolize access to material resources, claims another materialist theory, men garner higher status than women \cite{Sacks1974}, \cite{Friedl}. Expectation states theory argues that when some characteristic that is distributed among humans comes to be widely admired (e.g., agency in capitalist systems), then people who are expected to have that characteristic (e.g., men) will gain prestige and power \cite{Ridgeway2004}. Similarly, gender role theory argues that stereotypes of men and women derive from the social roles they occupy, respectively \cite{Conway}, \cite {EaglySteffen}. In turn, people use gender stereotypes to give or exclude people from particular roles, such as jobs \cite{Wood}, \cite{Glick1988}.

All of these theories posit an interplay among individual, group, and societal levels of interactions that are bidirectional. Further, they suggest on-going history in which aspects of the past (e.g., the distribution of men and women into roles, the characteristics associated with prestige) influence the present. In other words, micro-processes that occur between individuals produce or reproduce gender-based inequality. This suggests that the development of gender (and other group-based) inequality can be modelled as a dynamic system, wherein multiple actors potentially influence all the others, and may produce large-scale outcomes that no one necessarily intended.

In fact, Social Dominance Theory (SDT) postulates that group-based dominance hierarchies are complex dynamic systems that must be understood by considering the relationships between different levels of analysis, from individuals to families to groups to societies and beyond \cite{PrattoEtOthers2013}. SDT specifies how processes at one level of analysis influence processes at other levels. For example, people who favor group dominance support political actors, practices, and institutional policies that favor dominant groups \cite{PrattoEtAl1997B}. institutions tend to select and promote individuals whose values and psychological orientations concur with the institution’s role in sustaining or mitigating group hierarchy \cite{PrattoStallworth1997},  \cite{Haley2005}. Robustly across nations, people in dominant groups favor group dominance more than people in subordinated groups do \cite{Lee2011}. In fact, the gender difference in favoring general group dominance, called social dominance orientation, is larger in more individualistic and egalitarian societies in which those values are salient \cite{Lee2011}. Societies have feedback loops, then, reinforcing the way individuals relate to one another directly, and through their cultural ideals and institutions \cite{Pratto1999}. The kinds of empirical studies cited show pieces of the dynamic system, but only in brief “film clips” of processes, or “snapshots” of them. Research that tests dynamics in action across different scales is still needed.

The present research is the first dynamic test of how gender inequality might develop considering that there is no initial gender discrimination. Clearly there is nothing inherently superior or inferior about either women or men, so our model is not based on different abilities or survival advantages. Nonetheless, the way women and men are socialized and the roles they play provide some different learning experiences. To be parsimonious, we attempted to identify just one feature that could differ between simulated men and women that could produce gender inequality because of how individuals interact with one another.

We start with a human universal: the most fundamental social problem that all humans face is to belong to some collective of others \cite{BaumeisterLeary}, which is necessary for survival and thriving \cite{Caporael1996}. In fact, the sociometer theory posits that self-esteem is a tool for measuring how well one is accepted by others \cite{Leary2005}, \cite{Learyetal1995}. We therefore chose social esteem, both for self and others, as the causal parameter for our model.

Self-esteem can have many sources. One may view oneself as having virtue according to the cultural worldview of one’s society, e.g., \cite{Wisman}. A similar source is identifying as a member of a well-regarded social group \cite{Turner1975, turner1986social}. These sources of self-esteem hinge on having a socially-embedded (interdependent) self-concept that unites one’s own identity with other people \cite{Markus1989}. A separate method of obtaining self-esteem is to compare oneself favorably with others, either privately (social comparison) \cite{Festinger1954theory} or publicly, using status displays \cite{Mize}, \cite{Lee2015}. These latter methods require differentiating one’s self-concept from that of others, reflecting a more independent self-concept \cite{Markus1989}. There is the ample evidence that women are more communal than men, and that obtaining status is more valued by men, and that competing for status is prescribed for men by stereotypes \cite{SClingP1991}, \cite{Hoyenga}, \cite{Huberman}, \cite{Prentice}.  As such, we posit that men care more about others’ status than women do regarding cues to their own self-esteem.

On that basis, our model considers a population of virtual agents who interact to update how high they hold others and themselves in esteem. Indeed, to build evidence for this multi-level, dynamic theory, one must connect studies on humans at one or two levels of social organization together (e.g. \cite{Pratto1999}). The dynamic tenets of SDT, such as historical change, must be tested using modelling (e.g. \cite{PrattoStewartEtAl2013}), but this has not yet been done. Likewise, a glass ceiling effect cannot be diagnosed only from a single measure at a given time because the phenomena is about career trajectories \cite{RoseHartmann}. For these reasons, we use agent-based modelling to understand how the status structure of a society with two kinds of agents develops. Agent-based simulation is an appropriate tool to study complex interactions among people, groups, and society, because it allows one to observe the evolution of virtual agents who communicate amongst themselves \cite{Smith2007}, \cite{Mason2007}, \cite{Hedstrom2010}.

Further, we built the hypotheses of our updating function for the evolution of esteems on cognitive psychology studies. All societies construct categories of gender and the meanings of those identities \cite{heritier1996masculin}. In individualistic societies, there is a high degree of similarity in gender stereotypes across cultures \cite{WilliamsBest}, \cite{CuddyEtAl2015}. Very early, children learn gender roles \cite{Albert1988}, \cite{durkheim1922education}, namely the set of norms, cognitive and social behaviors described by the stereotypes of girls and boys, women and men \cite{henslin1999becoming}, \cite{belotti1991dalla}. The sanctions that children and adults receive for violating stereotypes of the gender others presume them to have provide disincentives for not conforming, and so, gender identity can become an important part of people’s self-esteem, especially for men \cite{Bosson}, \cite{Huberman}, \cite{Mize}, \cite{Prentice}. The aspects of socialized gender differences we posit as especially germane to self-esteem are (a) having a communal, interdependent orientation, or (b) being individualistic and sensitive to one’s status rank with respect to others \cite{bakan1966duality}, \cite{Carli2013}, \cite{Hall1999}, \cite{Abele2007}.

Further, there are cognitive gender differences that support our supposition, especially in how men and women vary in processing information about people.

\begin{itemize}
	\item women have higher attributional complexity (the degree to which an individual considers multiple kinds of information regarding someone in deep processing for social judgements) than men do \cite{Foels2010}, \cite{Westetal};
	\item concerning information processing, women (more than men) are concerned of others, have a lower threshold for message elaboration and extensively use message cues, all indicating they generally process more messages from others and rely less on heuristics to decide the value of a message \cite{Meyers-Levy1988}, \cite{Meyers-Levy1991}, \cite{Meyers-Levy2015}, \cite{Darley1995}, \cite{Vilela2016}, \cite{Putrevu2001}, \cite{Kempf2006};
\end{itemize}

Moreover, according to the Elaboration Likelihood Model of \cite{PettyCacciopo1986}, people who are less motivated to elaborate on the message use external cues to decide whether to accept  the message. We posit that such a cue can be the status of the person who conveys the message. As such, one would expect men to rely on status cues as a heuristic more strongly than women do, when they have to deal with information related to people.

\begin{figure}[ht]
\centering
\includegraphics[width=0.7\linewidth]{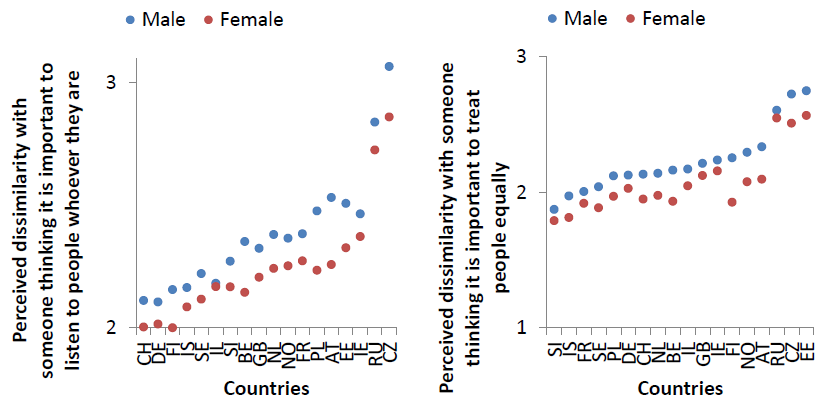}
\caption{\small{Perceived dissimilarity with someone thinking that it is important to listen to people however, for different sexes and in different countries. Perceived dissimilarity with someone thinking that it is important to treat people equally. Source: ESS8-2016 data, ed.1.0.}}
\label{data}
\end{figure}

These differences in how women and men respond to others and judge them are observable in results from several surveys. For example in Fig. ~\ref{data} we can observe that in all the European countries, women are more concerned to treat others equally, and are more open-minded in listening to others, than men are.

Uniting this gender difference in information processing with the gender difference in the importance of social status, we hypothesize that when exchanging information about people with other people:

\begin{itemize}
\item women (compared to men) will more influenced by the contents of the message itself, especially by how it is different from what they already know -- and this influence will be less contingent on their status evaluation of the message-giver than is the case for men;
\item status of self and others is salient to men, and so for men, the importance of the message content depends on how they perceive the message-giver’s status to be relative to their evaluation of their own status.
\end{itemize}

We implement these complementary hypotheses using an agent-based simulation that models the dynamics of interactions between persons (agents) over time, assuming that the population includes two subsamples who differ in their sensitivity or indifference to self-other status differentials. We assume these hypotheses are sufficient to explain the genesis and the maintenance of gender inequality. A brief introduction to the agent-based model and the motivation for our modelling choices will be given in the Methods section.

The agent-based model is used to examine the development of inequality between “male-like” and “female-like” agents, who merely differ in their sensitivity to status, and further, to test whether their interactions produces a status hierarchy, and one theoretical prescribed by SDT in that it is stable, with men having more status than women. Specifically, SDT has a tenet concerning the disproportionate rank of individuals in societies as a function of their gender. The “iron law of andrarchy” states that the higher in rank, authority, power, or status an individual is, the less likely that individual is to be a woman \cite{SidaniusPratto1999}. One version of this disproportionality, discovered independently, is the “Glass Ceiling,” the institutional barriers present even in supposedly non-discriminatory settings, that prevent most members of subordinated groups such as women, racial minorities or working-class people, from rising to the upper rungs of the institutional hierarchy, regardless of their qualifications or achievements. This concept was popularized in a prominent \textit{Wall Street Journal} article by Carol Hymowitz and Timothy D. Schellhardt, in 1986, entitled: "The Glass Ceiling: Why Women Can't Seem to Break the Invisible Barrier That Blocks Them From the Top Jobs."  For recent empirical evidence of glass ceilings, see \cite{Acker2007}, \cite{Folke}, \cite{Wright}. A related phenomenon is the persistence of salary and wage gaps that favor men over time. Longitudinal studies on labour data have documented this effect in several countries \cite{Arulampalam2007}, \cite{Albrecht2003}, \cite{Nordman2009}.

To sum-up, starting from the large literature on gender differences, our agent-based model assumes that women are more open to listening to others’ opinions than men are, whereas men more consider the relative status of others in determining how much to take the other’s views seriously with respect to social esteem.
 
\section*{Results of simulations}
This section shows that our agent-based model produces an emergent intergroup inequality that has the form of a glass ceiling inequality and that is not related to minority/majority effects. The indicators and the methods for their analysis are detailed in the Method section.

\subsection*{The emergence and the persistence of inter-group inequality}

\begin{figure}[ht]
\centering
\includegraphics[scale=0.35]{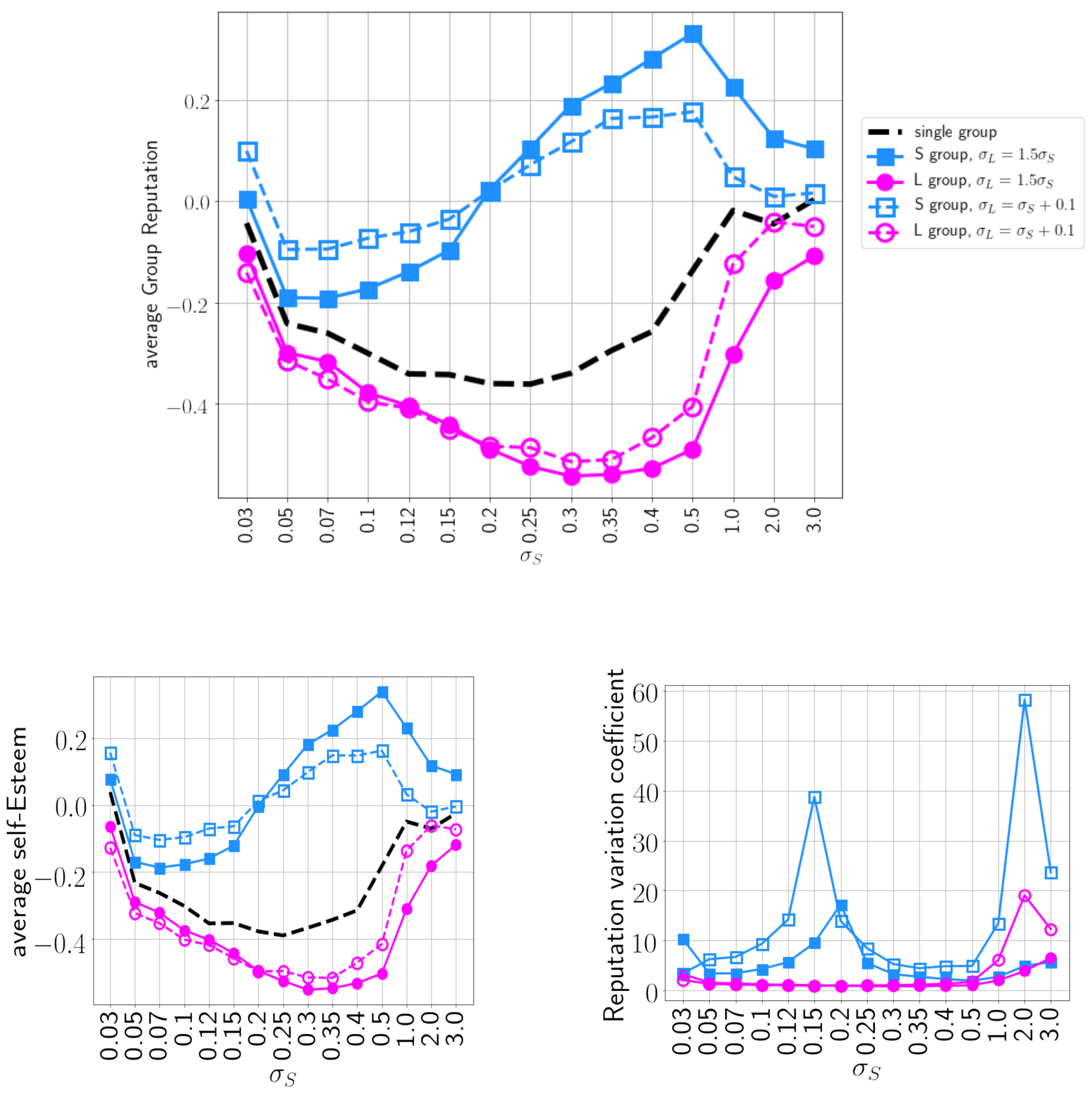}

\caption{\small{Upper plot: average group reputation for different values of $\sigma_S$. Lower left plot: average self-esteem of the groups for different values of $\sigma_S$. Lower right plot: Intra-group variability - variation coefficient for different values of $\sigma_S$. The black baseline represents a single group with open--mindedness parameter $\sigma_S$. In the other cases the population is divided into two equal-size groups with open--mindedness parameter $\sigma_S$ (blue lines) and $\sigma_L=f(\sigma_S)$ (magenta lines). The empty symbols correspond to the function: $\sigma_L=\sigma_S+0.1$ the others to the case $\sigma_L=1.5xsigma_S$}}.
\label{ineq}
\end{figure}

To better measure the effects on the dynamic interactions of two sets of agents, we compare the outcome of our model with a baseline simulation that has the same number of agents, but in which all the agents have the same open-mindedness levels as that of the lower-level group in our model. The black line in Fig. \ref{ineq}, which represents the baseline model, show that different values of the open--mindedness parameter $\sigma_S$ produce different average group reputation (i.e. average of esteems for the members of a group) and average agent self-esteem. These baseline plots are a summary of the model analysis presented in \cite{Huet2017}. The fact that both the average reputation and average self-esteem are negative is due to the model settings about the gossip dynamics in the interactions \cite{deffuant2018dark}. %%In the SI we show that the results for the two groups, in comparison with the single group baseline, are stable under different model conditions.%% 

Comparing the two-group model with the baseline model, we immediately observe a differentiation of the two groups’ behaviors: the reputation of the group with lower open--mindedness (the S--group) is always above the baseline (see the upper plot of Fig.~\ref{ineq}). The lower open—mindedness (“male-like”) group, on average, gains large advantages by interacting with the rest of the population.

Moreover, the average reputation of the S--group is not only higher than agents’ in the baseline model, but also higher than the average reputation of the higher open--mindedness (“female-like”) group (the L--Group). We can therefore conclude that this status-formation process, where the two sets of agents become differentiated in their reputations (public status) due only to their initial levels of open--mindedness, generates a dominant group with an average status reputation higher than the second, disfavoured, but more open-minded group. One could describe the result, alternatively, as showing that responding more to those equal or superior in status to oneself results in attaining higher status.

The same results just described for reputation are also apparent for self-esteem; in the left lower plot of Fig.~\ref{ineq} we see that the less open--minded group exhibits higher self-esteem on average than both the baseline model and than the other group. 

\subsection*{The emergence of differences in intragroup stratification}
Finally, we analyze intra-group inequality in reputation, using the reputation variation coefficient. The right lower plot of Fig.~\ref{ineq}) shows that the dominant (less open--minded) group displays higher variability among those agents’ reputations than the variability in the other group. In other words, in addition to developing inter-group inequality in consensual reputations and in own self-esteem favoring the less open-minded group, the model also creates much stronger intragroup stratification in reputations of agents in the more dominant group, with very little status differentiation among the more open-minded group.

Note that none of the group difference effects shown in Fig. ~\ref{ineq}) change as a result of the level of lower open--mindedness parameter ($\sigma_L=f(\sigma_S)$).

\subsection*{Replication of the dynamics of glass ceiling}

Delving into the stratification process, we will now show that the model’s dynamic system, in addition to developing the group differences in self-esteem, status reputation, and level of stratification in reputation already shown, produces status inequality that has the same static and dynamic characteristics of the glass ceiling, as defined by  \cite{cotter2001glass}.

\begin{figure}
\centering
\includegraphics[scale=1.1]{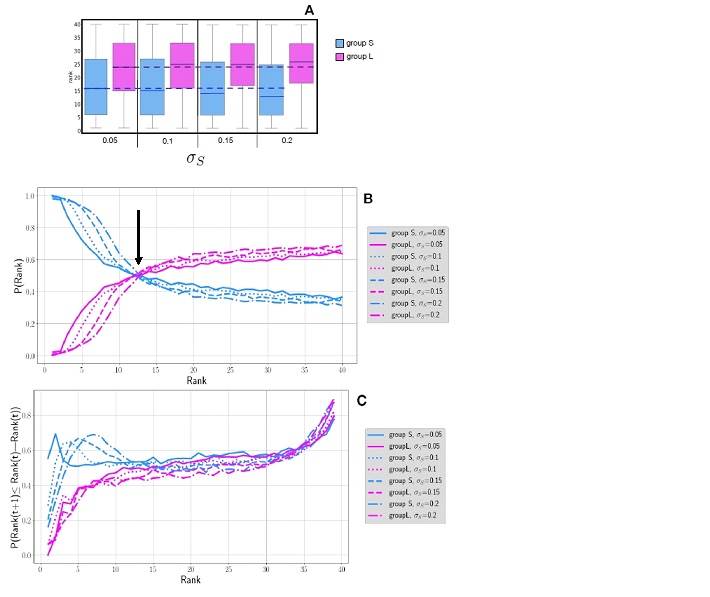}
\caption{\small{Group S (in magenta) is less open-minded than Group L (in blue), and then occupies on average higher ranks (1 is the best rank). Plot A: Boxplot of status rank values for different populations having different values of the open--mindedness parameter $\sigma_S$ for the group S, and the corresponding value of ($\sigma_L$ = 1.5 x $\sigma_S$) for the group L. Dashed lines indicate group means for the lowest $\sigma_S$ 0.05, across levels of $\sigma_S$: we observe the difference of average ranking for group S and group L increases with the value of $\sigma_S$. Plot B: Probability for a rank to be occupied by an agent in a group at each level of the open-mindedness parameters (see legend, still with $\sigma_L$ = 1.5 x $\sigma_S$). Plot C: Probability of moving to a lower status position (higher in rank) given the agent’s initial rank, by group and open-mindedness parameter. The three plots represent averages computed from 100 replicas of each tested parameter set.}}
\label{glass}
\end{figure}

The first criterion given by \cite{cotter2001glass} is: \emph{A glass ceiling inequality represents a gender or racial difference that is not explained by other job-relevant characteristics of the employee}. In our agent-based model, agents have no job-relevant (or any other kind of) characteristics; they only differ by their open-mindedness or degree of insensitivity, assuming this is a cultural characteristic of their group (however defined: gender, race, etc.). Our model consistently produces two sets of agents with unequal reputations, where the less open-minded group becomes superior, a result that cannot be attributed to any job-relevant (or other agent-relevant) characteristic. This result is shown in Fig.~\ref{glass}A, which depicts boxplots for the rankings of both groups, varying over a range of the more open group’s open-mindedness levels. In Figure 3A, the group S with lower open--mindedness, alway occupies the higher status rank positions (scale with 1 being the highest rank, 40 the lowest), regardless of the level of the associated parameter $\sigma_S$.

The second criterion given by \cite{cotter2001glass} is: \emph{A glass ceiling inequality represents a gender or racial difference that is greater at higher levels of an outcome than at lower levels of an outcome}. Our model would reflect the second criteria if the frequency of the lower-reputation group agents (the group with higher open--mindedness, $L-Group$) decreases as the social rank improves (indicated by low order of prestige). Fig.~\ref{glass}B shows that the probability that a rank is occupied by an agent of the $L(S)-Group$, declines markedly among the highest social rank positions, which of course are occupied by Group S. This is not a linear function, however. Rather, there is a “glass ceiling” around rank 12 above which no agents of Group L rise, indicated with the arrow in Fig.~\ref{glass}B. Just below the elite ranks (about 1 to 12), there is a near-even chance those positions are occupied by Group L or Group S agents, as if a few women can have higher-than average rank, but still not rise to the very top. However, the entire bottom half of the social ranks (21 to 40) is more likely to be occupied by agents of the more open-minded than the more closed-minded group. Notice that the result is stable for all the values of the open--mindedness parameter $\sigma_S$.

The third and the fourth criteria given by \cite{cotter2001glass} are: \emph{A glass ceiling inequality represents a gender or racial inequality in the chances of advancement into higher levels, not merely the proportions of each gender or race currently at those higher levels} and
\emph{A glass ceiling inequality represents a gender or racial inequality that increases over the course of a career.}

Evidence for these criteria in our model would not just be shown by the probability of each group at each rank, as shown in Fig.~\ref{glass}B, but by observing that the probability for agents of the lower-prestige group to move to a higher rank is smaller than the probability that agents of the higher rank do. In other words, if our model produces a glass ceiling, not only should disfavoured group agents should be less numerous in higher ranks at a given time point, but also, when agents change rank in time, the disfavoured group ($L-Group$) should have a lower chance, compared to agents of the dominant group, to reach a higher rank.

This dynamical pattern is reproduced by our model. It is shown in Fig.~\ref{glass}C, where we compare the probability of obtaining a higher status at time $t+1$ starting from an initial position at time $t$: $P(Rank(t+1)\leq Rank(t)|Rank(t))$. We see that agents’ progression to higher ranks is less probable for the more open—minded, lower status group, especially in the higher positions.

\begin{figure}[ht]
\centering
\includegraphics[scale=0.35]{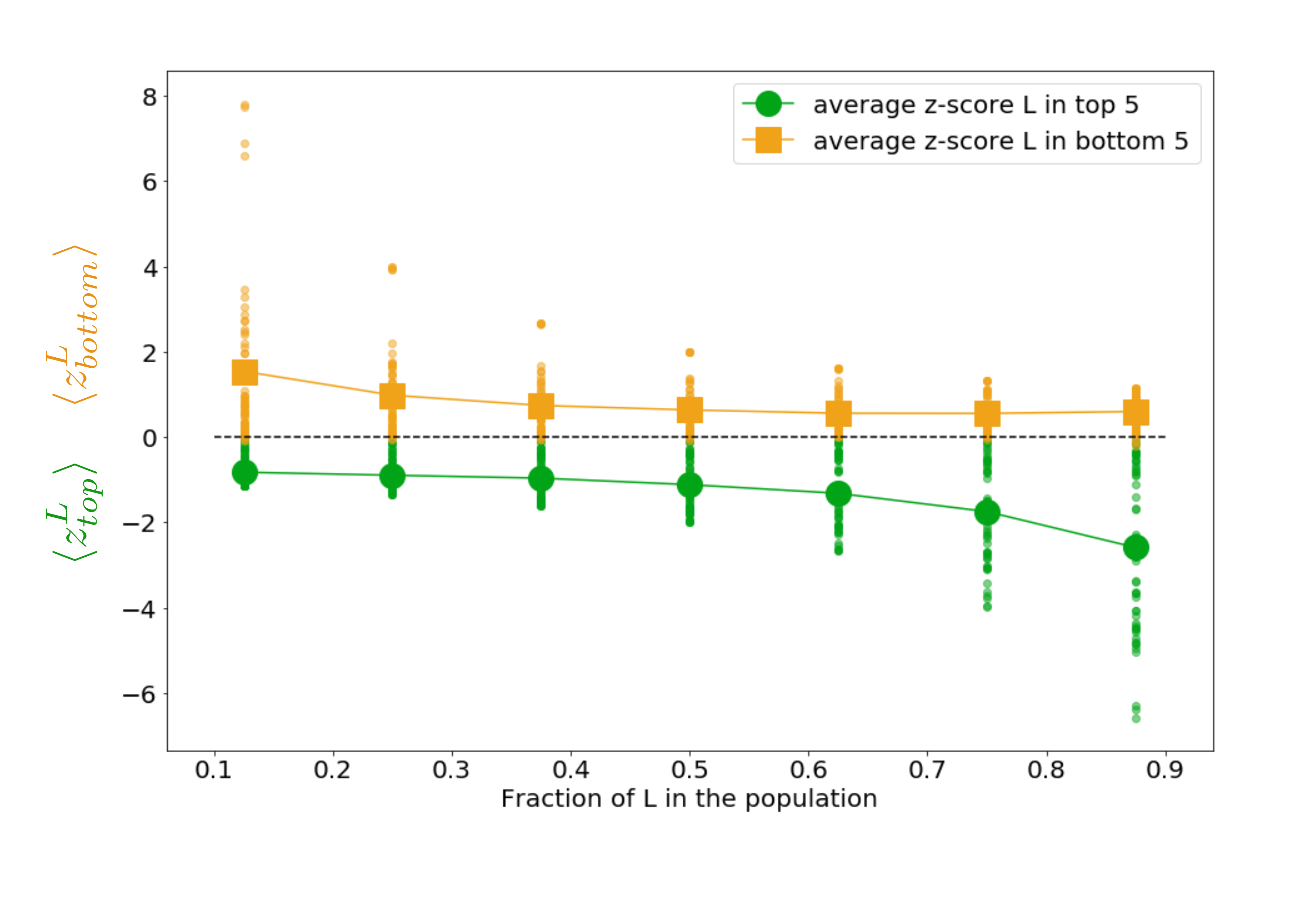}
\caption{\small{Z-scores, indicating how different from the expected probability from the fraction of L in the population, is the probability of the more open-minded Group L being in the 5 most prestigious ranks (green points, lower than the expected z-score 0) and of being in the 5 least prestigious ranks (orange points, higher from the expected z-score 0) plotted by their proportion of the population of agents. Each small point represents a simulation setup replicated 50 times, varying the combined values of ($\sigma_S$, ($\sigma_L$, $k$) - see the method section for details.}}
\label{size}
\end{figure}

\subsection*{Group size does not change the basic inequality patterns}

In the previous paragraphs we considered a population equally partitioned in the two groups of 20 agents each. We now investigate whether the relative size of the groups can change the stratification patterns previously observed (still with a total population of 40 agents). We elaborate a large experimental design described with more details in the method section. We vary by step of 5\% the fraction of L in the population. For each fraction, we varied values of $\sigma_S$ and $\sigma_L$, as well as the level of gossip through the parameter $k$ over a large range. Overall, we ran 405 various parameter sets replicated 30 times each. Considering the results of each set of experiments obtained for each fractional value of L, we compute the Z-score indicating how the results vary from the expected value computed from the fraction of L. Indeed, in an ideal case, where no inequality is present, the expected presence of the disfavoured group L in the top 5 and bottom 5 ranks should be proportional to their relative size in the population: $P^{exp}_{top5}=P^{exp}_{bottom5}=\sim N_L/N$. Therefore, in Fig. \ref{size} we should observe all the points distributed around the dotted black line 0. On the contrary we observe that, for all values of the simulation parameters, and for all the values of the relative group size, the presence of L-Group’s (female-like) agents in the top 5 ranks is lower than chance, but higher than chance in the bottom 5 ranks. 

This result shows that the mechanism of stratification due to behavioral difference in our model is valid not only for equal sized groups, but also for populations with minorities.

\section*{Discussion}
This paper presents an extremely parsimonious but rich experiment to test one simple hypothesis of the genesis and development of gender inequality. Drawing on the basic need to belong, aspects of gender socialization regarding self and others, and cognitive gender differences, our model posited one gendered behavior concerning listening and information-processing, namely, that men’s estimations of self and others would depend on the others’ relative status rank more than women’s would, because women respond to other’s messages more, with less regard for the status of the other. The model represents this difference solely as the value of the open-mindedness parameter for each set of agents. We showed that this cognitive behavioral difference generates inter-group status inequalities and additional features of status inequalities that are often found in human societies. 

The model evolution shows that:

\begin{itemize}
	\item The average reputation of male-like (more status-conscious) agents is generally higher than the average reputation of female-like (more open-minded) agents. Thus, the model replicated the near-universal finding that men have higher public prestige than women do across cultures, according to anthropologists \cite{Rosaldo}, \cite{Sanday}, and to lay people \cite{WilliamsBest}, \cite{CuddyEtAl2015}, \cite{ConwayVartanian}. Further, like the many studies that show that there is a high degree of consensus between men and women that men have higher status \cite{Glick2000}, our model showed that male-like and female-like agents were in accord as to which of the agents had higher and lower reputations. 
	\item The average self-esteem of male-like agents is higher than the average self-esteem of female-like agents. This replicates several empirical studies comparing women and men, especially \cite{Bleidorn2016};
	\item There is more status differentiation among the more status-conscious, male-like agents, than among the more open-minded agents, which replicates the broad finding that most human societies, most public status-ranking occurs among men \cite{Rosaldo}, \cite{Sanday}.
	\item When male-like and female-like agents interact, the male-like agents’ reputations increase, but the female-like agents’ reputations decrease. This “gender” differentiation as a function of interactions among agents parallels the finding that people tend to emphasize their gender-stereotypic features when they interact in a mixed-gender context \cite{GuimondEtaL2006}. In fact, prescribed gender roles suggest that women should be modest, whereas men can be proud and arrogant \cite{Prentice}. People’s genders are more salient to them in mixed-gender settings \cite{MacGuire1976}.
	\item Female-like agents are more subject to peer influence than male-like agents are because female-like agents have lower average self-esteems. This is in accordance with data on humans \cite{eagly1978sex} and \cite{Carli2001}.
	\item Glass ceiling \cite{cotter2001glass}: the highest ranked positions are rarely reached by female-like agents, and the chances that a female-like agent advances her status rank in the future is smaller than the chance of status rank advancement for male-like agents.
\end{itemize}

We can conclude that the present simple initial group difference in open-mindedness converges on a somewhat consensual status-system in which the more open, female-like agents develop lower self-esteem and lower standing and prestige in the eyes of other agents, while the male-like agents develop higher self-esteem, more status differentiation among them, as suits their “concern” with status, and lower regard for female-like agents.  

In addition to the parallels between our model‘s results and empirical findings concerning gender status inequality mentioned above, the present results have a number of implications for theories of gender inequality and other theories of group-based inequality.

Our model had no separate gender roles, no initial difference in gendered status, no kind of gender discrimination or prejudice, no family obligations, no stereotypes, no authorities, no violence, no division of labor, no patriarchal religion, no accumulative economy, and no inheritance or marriage systems, all of are known to contribute to gender inequality \cite{Friedl}, \cite{Kleinjans}, \cite{Rosaldo}, \cite{Schwendinger}, \cite{Sanday}, \cite{TreasTai}. Yet the results showed patterns that mimic real gender inequality in several details: The model never produced a “societal structure” in which the more open-minded (“female-like”) agents had higher self-esteem nor higher prestige than the more close-minded (“male-like”) agents. The gender inequality in self-esteem finding reflects robust empirical findings \cite{Bleidorn2016} as does the gender inequality in public prestige \cite{Rosaldo}. The model showed that even when any more open-minded agents attained above-average ranks, there was a structural barrier above which nearly none of them ever attained the top few ranks. This glass ceiling effect \cite{cotter2001glass} reflects nearly every contemporary society, where some women can become “middle managers” or attain authority roles (where they will be a minority), but where women in as chief executives in business \cite{Adams}, \cite{Yamak}, government, religious institutions \cite{Murphy}, \cite{Oluwaniyi}, \cite{Rasool}, educational institutions \cite{Davies}, \cite{Wong}, \cite{Kalaitzi}, or are the leading intellectuals or artists, are singular exceptions \cite{Tate}, \cite{Lincoln}, \cite{Hargittai}. The model also showed that for the bottom half of the prestige structure, a relatively small, 60\% or so, majority of agents were ``female-like''. The association of gender with prestige rank is precisely what makes gender a status-characteristic \cite {Cejka}, \cite{Riordan1983}, \cite{Ridgeway2004}. 

Further, the model showed that the stratification among male-like agents was much stronger than the stratification among female-like agents. This finding mimics the fact that status does differentiate men more than it does women in many societies \cite{Hoyenga}, \cite{PrattoWalker}. The insight our model provides to findings that men are more competitive and want to pursue public status more than women do on average \cite{Lee2015}, \cite{Mize} is that the Leviathan process itself, with agents who differ in their open-mindedness to being influenced is sufficient to produce this effect. As with any experiment, ours cannot prove that the process the experiment showed evidence for what occurs outside the experiment, but our results do provide an extremely parsimonious process by which the real-world gendered phenomena occur. 

The fact that our experiment reproduced these four findings about gender inequality and stratification raises the question about whether alternative theories about gender inequality and stratification are incorrect or unnecessary. We would argue against the idea that our model renders gender theories unnecessary for several reasons. First, models are but one research tool, and as with any experiment, our model serves as an existence proof of a process that can account for “naturally” (or socially) observed phenomena. No experiment can prove that the process that occurs within the experiment is the process observed outside the experiment. Second, we derived the causal parameter of the model based on empirical research on humans. The progress of science entails an ongoing dialogue among results from several research methods, in the present case, where empirical research on humans informs the model, and then, how the model may inform theories about humans. 

So let us turn to the question of where our model fits with respect to theories of gender inequality on humans. The first point is that our model examined a “micro” set of processes over time, that is, solely concerning interactions among agents, who can be thought of as individual persons, yet it produced realistic “societal” patterns without any goals or intentions about those patterns in the agents themselves. In other words, processes at one level of social organization can indeed produce effects at a higher level of social organization, as SDT (Social Dominance Theory) holds. This demonstrates that one does not need to assume that any particular person intends for a societal outcome, or even anticipate the outcome that their actions help to produce. Considering SDT in particular, substantial evidence shows that people who are higher rather than lower on social dominance orientation do in fact reinforce and even strengthen group-based inequality, by selecting roles in institutions that serve those ends, \cite{Haley} for example. There is also evidence that people higher on social dominance orientation tend to believe that the social world is a zero-sum game for social status. That is, they believe that if one does not gain superior status, they will become inferior. Potentially, one of the motivations for the many of the actions higher SDO people that do maintain group-based hierarchy is simply to show others that they understand the hierarchy and are following its norms. Showing that one knows one’s cultures norms and conforming to them is a universal way of ensuring acceptance by others. Alternatively or in addition, higher SDO people may be motivated to attain higher status, and perhaps also to ensure that that status means something because they do in fact belong to the highest groups in group-based status hierarchies. This interpretation is consistent with there facts: (a) SDO among higher-status groups is positively associated with their differentially identifying with high- but not low-status groups \cite{PrattoStewart2012}, (b) SDO has a mild association with status-striving \cite{PrattoEtAl1997B},  and (c) SDO is higher among higher-status than lower-status groups \cite{Lee2011}. The fact that our model was based on a single motivation and micro-processes stimulates generating micro-processes related to SDO other than that they want specifically to maintain a group-based dominance society. They confirm, also, that the very existence of individual and group variation on SDO cannot account for the existence of group-based dominance hierarchies \cite{PrattoEtOthers2013}.

In fact, it may be useful to consider our model as representing a micro-process that is part of a broader context, ones we may have replicated, but did neither measure nor manipulate.  For example, regarding our finding of greater status reputation stratification among the male-like than the female-like agents, we can consider additional reasons for the difference between the agents’ status sensitivity. Certainly some aspect of gender role socialization leads to the difference in open-mindedness we modelled, and there are many processes that could produce the modelled difference. At least in capitalist and individualist cultures, boys are born into a society where competitiveness among boys, and the stratification among men means that they could either do well or do very poorly. To survive these difficult contexts for maintaining positive self-esteem \cite{VandelloEtAl2008}, males may protect themselves by disregarding anyone’s opinion of them which is lower than their self-esteem (the Leviathan process), and attend only to the opinions of those they “have to” -- namely higher status people. Further, as members of both sexes observe that males have higher status than females, all of them learn the gendered status structure, but for males, endorsing that hierarchy (and hostile sexism) is linked to their striving for positive self-esteem. Men may seek higher-prestige occupations than women do due not just to fulfill gender role stereotypes, but to have another basis for demonstrating their status value, \cite{Kuchynka} namely, having a high-status occupation \cite{Buss1993}, \cite{Kleinjans}. In fact, a number of theories of the relation between sexism and masculinity argue that self-motivations are involved \cite{Bosson}, \cite{VandelloEtAl2008}, and those self-motivations make men especially aggressive against women \cite{VandelloEtAl2008}, \cite{Dahl}, \cite{WeaverEtAl2010} and endorse sexism and sexist political parties \cite{DiMucci}, \cite{GBS}.

Finally, one point related to the model has to be pointed out. Our groups, composed of agents having a different open--mindedness, communicate with each other intensively: indeed an agent from one group has the same probability to communicate with an agent from its group than an agent from another group. We know such an hypothesis is far from the reality since prejudiced people avoid intergroup contact \cite{Pettigrew1998}. Then the next step for this research is to study different protocols of communications between groups, particularly to see the impact of different communication patterns on the inequality.
 
\section*{Method}

In this section, we present at first an overview of the mechanism for status construction of the model. Then, we describe in more details the dynamics of the evolution of agents of our study. The following subsections present the experimental designs realised to obtain our present results, as well as the indicators extracted from the measures of the simulated dynamics, and how they have been analyzed.

The last subsection makes a short review of the previous studies of the macro behaviours emerging from this model, and explains how the results of the present study can be explained from the agent-based model point of view.

\subsection*{The social mechanisms for status construction in the agent-based model}

A person’s status is the relative prestige accorded to the person in a society. From the dynamical point of view, status stratification can be described as the emergent (static or dynamical) equilibrium of a system of exchanges among the actors about themselves and others. A large literature indicates that these status stratification processes could be a priori connected to cumulative advantage mechanisms \cite{merton1968matthew}, \cite{merton1988matthew} - also defined as Matthew effects or Matilda effect \cite{Rossiter1993}. The effect of cumulative advantage as generative mechanism of status stratification has been analyzed in several papers \cite{lynn2009sociological}, \cite{gould2002origins}. This has been combined with deference exchange mechanisms in prior modelling research in \cite{manzo2015heuristics}.

Differently from such research, the Leviathan model, presented by \cite{deffuant2013}, implements a less explicit form of the Matthew effect cumulative advantage process. Leviathan presumes that actors may not be aware of the consensual status of their peers, but can, however, compare their own self-esteem with the esteem they have for each peer. We will use the framework of the Leviathan model \cite{deffuant2013}, and more especially its simplified version \cite{Huet2017} to study the evolution of the statuses of the actors, assigning to each actor one of the two cognitive traits described above that are associated with gender.

The Leviathan computer model considers a population of virtual agents, each of whom has an esteem about each of the agents (including itself). These esteems change, for each agent, via random dyadic encounters during which two agents “talk” about each other - and also “gossip” about other agents. The influence mechanism resulting from these encounters can change each agent’s self-esteem and each agent’s esteem of the others. To be realistic, their communication is not perfect: a noise parameter slightly and randomly changes what is “said” compared to what is “heard.” Initially, all agents hold neutral estimations of themselves and each other agent.

In human research on opinion influence, there are two main social comparison factors. One is similarity between the agents, or homophily \cite{axelrod1997dissemination}, \cite{byrne1997overview}, \cite{mark2003culture}, \cite{takacs2016discrepancy} (see \cite{castellano2009statistical}, \cite{flache2017models} for a review), in which one’s influence on another increases with similarity to the other. The other factor that increases influence, which is employed in the Leviathan model, is the superiority of the speaker as perceived by the listener. This factor complements homophily effects because a large body of work finds that influence is largely based on the credibility of the source \cite{crano1972preliminary}, \cite{pornpitakpan2004persuasiveness}, \cite{tormala2007attitude}.

To be specific, in our model, the influence function is a classical smooth threshold function \cite{huet2007taking}, \cite{young2009innovation} of the difference between the listener’s self-esteem and her esteem about the speaker. This difference can be viewed as the agent’s perceived relative status. The smoothness of this function is driven by an open--mindedness parameter, $\sigma$, that defines how strongly agents rely on their perceived hierarchy among agents to define someone else’s credibility. The higher the open--mindedness, the less sensitive the agent is to the perceived position of other agents in gauging the agents’ credibility. In other words, the higher one’s open--mindedness, the more equally influential the other agents are on one. Prior research on this model has shown that, in a totally homogeneous population of agents having the same open--mindedness, various types of hierarchies emerge and persist according to various simulation parameters (see Methods for more details).

Based on the considerations in the preceding paragraph, we designed for our present study a Leviathan model in which  the agents are divided in two sets, one set, or “group” characterized by a small value of the open--mindedness parameter ($\sigma_S$), which our arguments above suggest more characterizes men, and another group with a larger value ($\sigma_L$), representing women. We test the hypothesis that this behavioral difference among agents is sufficient to generate intergroup (gender) inequalities in terms of the average esteem the population has for each agent in each group. Note that these predictions are not contingent either on intergroup prejudice and discrimination, nor on agents categorizing the other agents as members of the groups (for such models, see \cite{Uhlman}). Further, we predict that the inequality that will emerge from our model will do so regardless of other parameter choices.

\subsection*{The details of the influence-Leviathan model}

In the Leviathan model each agent $i$ is endowed with:

\begin{itemize}
\item a self esteem $a_{ii}(t)$
\item a set of esteems about the other agents $a_{ij}(t)$
\end{itemize}

We initialize the model with a set of $N$ agents having all the self-esteems and all the esteems on the others set to zero: $a_{ij}(0)=0, \forall (i,j) \in [1,...,N]\times[1,...N] $.

The individuals interact in uniformly and randomly drawn pairs $(i, j)$ and, at each encounter, they try to influence each other on their respective values. We define one iteration, i.e. one time step as $N/2$ random pair interactions (each individual interacts 1 time on average during one iteration). To be more precise, one iteration involves $N/2$ executions of the following steps:

\begin{center}
\fbox{\begin{minipage}{25em}
\begin{itemize}
	\item Choose randomly an agent $i \in [1,...,N]$
	\item Choose randomly another agent $j$ in the same group than $i$ with the probability 0.5, else in the other group
	\item $Influence(i,j)$
	\item $Influence(j,i)$
	\item $Gossip(i,j)$
	\item $Gossip(j,i)$
\end{itemize}
\end{minipage}}
\end{center}

Each dyadic interaction involves two different processes: $Influence$ and $Gossip$. The $Influence(i,j)$  process defines how the agent $i$ changes her esteem on himself and on $j$, according to $j$’s influence.  The gossip process $Gossip(i,j)$ defines how agent $i$ changes her esteem on other $k$ agents, according to $j$’s esteem on these agents. The $Influence$ process is defined by the following equations:

\begin{center}
\fbox{\begin{minipage}{28em}
\begin{equation*}
    a_{ii}(t+1)=a_{ii}(t)+p_{ij}(t)\left[a_{ji}(t)-a_{ii}(t)+Random(-\delta,\delta)\right]
\end{equation*}
\begin{equation*}
    a_{ij}(t+1)=a_{ij}(t)+p_{ij}(t)\left[a_{jj}(t)-a_{ij}(t)+Random(-\delta,\delta)\right]
\end{equation*}
\end{minipage}}
\end{center}

The $Gossip(i,j)$ process is the following. During an encounter, we suppose that agent $j$ propagates to $i$ her esteems about $k$ agents randomly chosen in the population, according to the process:

\begin{center}
\fbox{\begin{minipage}{28em}
Repeat $k$ times:
\begin{itemize}
	\item Choose randomly $z\in[0,...,N] : z!={i,j}$
    \item$a_{iz}(t+1)=a_{iz}(t)+p_{ij}(t)\left[a_{jz}(t)-a_{iz}(t)+Random(-\delta,\delta)\right]$
\end{itemize}
 
\end{minipage}}
\end{center}
 
The strength of the propagation of esteem, in both the influence and the gossip processes is ruled the influence function $p_{ij}$ that implements the hypothesis that the more $i$ perceives $j$ as superior to herself, then the more $j$ is influential on $i$. It is a logistic function (with parameter $\sigma$) of the difference between the esteem of $i$ about $j$ ($a_{ij}$) and the esteem $i$ about self ($a_{ii}$):

\begin{equation*}
	p_{ij}(t)=\frac{1}{1+e^{-\left[ a_{ij}(t)-a_{ii}(t)\right]/\sigma}}
\end{equation*}

The influence function, $p_{ij}$, tends to $1$ when $a_{ij}-a_{ii}$ is close to the maximum possible distance between $i$ and $j$, namely when $i$ evaluates $j$ much more than himself. It tends to $0$ when $i$ values $j$ lower than herself. The open-mindedness parameter,$\sigma$ defines the slope of the function close to $a_{ij}-a_{ii}=0$.

\begin{figure}
\includegraphics[scale=0.7]{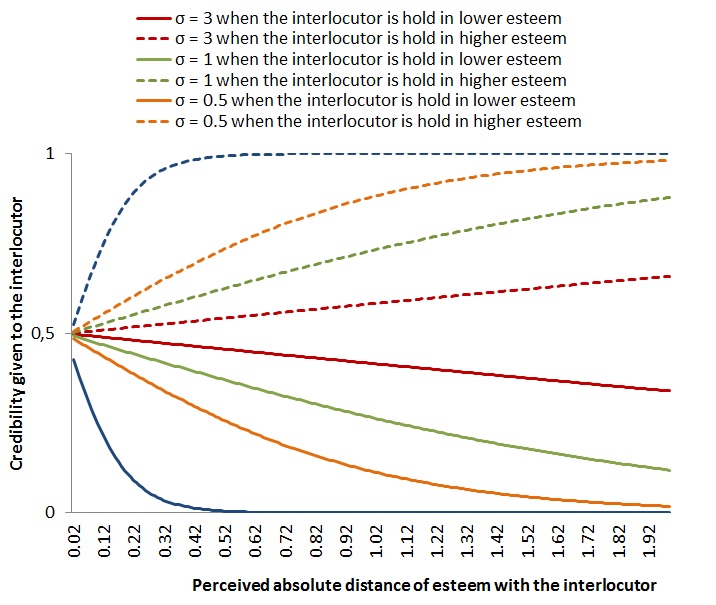}

\caption{\small{Examples of credibilities (i.e. strength of the influence) given by an agent $i$ to an agent interlocutor $j$ depending on the absolute perceived distance $i$ between $j$ and him-herself (on the x-axis), and the fact $i$ hold $j$ in lower esteem (plain lines) or higher esteem (dotted lines), for $\sigma$ = 0.1 (in blue), 0.5 (in orange), 1 (in green) and 3 (in red). We observe that the higher the perceived distance, the lower the credibility for $j$ when held in lower esteem, or the higher credibility for $j$ when held in higher esteem. Moreover, the larger $\sigma$, the smaller the difference of given credibility to $j$ if he/she is held in low compared to high esteem.}}
\label{credibilityFunction}
\end{figure}

Figure \ref{credibilityFunction} illustrates these tendencies. One can notice that when $\sigma$ increases, becoming 3 for example (in red), the difference of credibility $p_{ij}$ given to an agent $j$ holds by an agent $i$ in higher esteem (dotted lines) compared to credibility $p_{ij}$ given to an agent $j$ hold by an agent $i$ in lower esteem (plain lines) strongly decreases, but remains. This is why in our experimental design, we tested more small values$\sigma$ than larger values. 

The parameter $\delta$ models the idea that an agent $i$ has no direct access to the esteems of another one ($j$) and can misunderstand it. To take into account this difficulty, we consider the perception of the agent $i$ of the esteems of agent $j$ as the value $a_{j\bullet}$ plus a uniform noise drawn between -$\delta$ and $+\delta$. This random number corresponds to a systematic error that the agents make about the others’ esteems.

Finally, the previously presented model has 4 parameters (i.e. in this version, we keep the probability of $i$ to meet an $j$ of another group to 0.5):

\begin{itemize}
	\item the number of agents, $N$
	\item the open--mindedness parameter, that in the paper we associated to gender identities, $\sigma$
	\item the noise intensity $\delta$
	\item the number of agents an agent talked about during a meeting, $k$
\end{itemize}

\subsection*{The simulation setup}

We presume here that a population is composed of two different groups which are distinguished by their open--mindedness. The group size can vary. Each group $G$ is characterised by two values: the number of agents of the group, $N_G$, and the group’s open--mindedness $\sigma_G$. $G$ take the values $S$ or $L$, where

$S$ stands for the group with a Small open--mindedness, $\sigma_S$ and $L$ stands for the group with a Large one, $\sigma_L$. We vary the size of each group, maintaining a fixed total number of agents $N=N_S+N_L$. We implement two possible relationship functions between the $\sigma_S$: $f_1: \sigma_L=1.5\sigma_S$ and $f_1: \sigma_L=\sigma_S+0.1$.

We ran the model assuming $N = 40$. In the first part of the results we consider the case where the agents are equally divided into the groups ($N_S=N_D=20$). In the last part, to test whether having majorities or minorities of the more- and less open-minded groups changed the results, we tested the model, varying groups' sizes by 5 agents, $N_S=0,5,10,15,20,25,30,35,40$. We also vary $\sigma_S$: 0.03, 0.05,0.07, 0.1, 0.12, 0.2, 0.25, 0.3, 0.35, 0.4, 0.5, 1, 2 and 3. For each $\sigma_S$, two various $\sigma_L$ are tested according to the laws presented in the previous paragraph. Each simulation is replicated 50 times. The length of the simulations and the frequency of the measures depends on the values of $\sigma_S$, the lower the longer the simulations to ensure we capture indications about the properties of the stable dynamics. There lasts at minimum 500,000 iterations with a measure every 10,000 iterations.

We set the noise parameter $\delta=0.2$ and the gossip parameter $k=3$, except for the study varying the size of the groups presented in the previous paragraph for which $k$ varies from 0 to 3 by step 1. We can then ensure results are stable.

\subsection*{Indicators of the evolution and methodology for their analysis}

In the Results section, we explain how the simulation results verified our micro dynamic hypotheses. To enlighten our results, this subsection explains how various measures of inequality are derived and analyzed from the model results.

In the first part of the Results section, we focus on outcomes that are derived from various ways of aggregating the agents’ esteems for one another across the population. We define the reputation of an agent as the average of the esteem accorded to this agent by all the other agents. The measure of inter-group inequality in status is the variability among the average reputation across the population for each group of actors. To understand within-group status differentiation, we define a variation coefficient to index intra-group variability in self-esteem, namely, the standard deviation of the group’s agents’ self-esteems, divided by the group’s average self-esteem. Finally, we define the socially-consensual degree of intra-group status differentiation as the standard deviation of that group’s agents’ reputations.

In the second part we delve into the individual dynamics and show that the model generates and maintains a \emph{glass ceiling} effect, shown in the level of access the agents in each group have to higher status positions. We compare the structural results of our model’s individual dynamics with the properties of glass ceiling effect dynamics precisely defined by \cite{cotter2001glass} as follows: 

\begin{itemize}
	\item A glass ceiling inequality represents a gender or racial difference that is not explained by other job-relevant characteristics of the employee.
	\item A glass ceiling inequality represents a gender or racial [status] difference that is greater at higher levels of an outcome than at lower levels of an outcome.
	\item A glass ceiling inequality represents a gender or racial inequality in the chances of advancement into higher levels, not merely the proportions of each gender or race currently at those higher levels.
	\item A glass ceiling inequality represents a gender or racial inequality that increases over the course of one’s career.
\end{itemize}

The results section shows that difference in open--mindedness between men and women is sufficient to generate gender inequality in prestige, and shows the properties of the glass ceiling effect listed above. The social rank of any agent is the order of that agent's reputation, when agents’ reputations are ordered from best (highest) reputation (1) to worst (lowest) reputation (N). To test for the glass ceiling effect, as defined by the preceding criteria, we will analyze the agents' social rank at different time points of the model’s history. We will first analyze the average ranking position. Second, we will study the probability that a given rank position is occupied by an agent of the S--group (that has smaller open--mindedness, namely simulated men) or by an agent of the L--group (that has larger open--mindedness, namely simulated women). Finally, to test whether there is differential advancement of the two kinds of agents, we will analyze the probability, for an agent with a given social rank, to progress, in future time points, to a higher rank.

\subsection*{How to explain the observed behaviours of the agent-based model}
\subsubsection{Short state-of-the-art}
Prior research on this model has shown that, in a totally homogeneous population of agents having the same open--mindedness, various types of hierarchies emerge and persist according to other simulation parameters \cite{deffuant2013}. However, the basic Leviathan model \cite{deffuant2013} includes two micro-processes driving the change of esteem of agents: (1) vanity implies on the one hand, that one agent flattered by another rewards the other agent by increasing the esteem he/she has for her/him and on the other hand that one agent despised by another punishes him/her by decreasing her/his esteem for her/him ; (2) influence between two agents regards how high they hold themselves and others in esteem, based on the perceived difference of esteem between them.

In this paper, we only focus on the second micro-process, the influence, which has been studied by \cite{Huet2017} and has been shown sufficient to explain the emergence and the persistence of different social orders in the population. Indeed, from such a micro influence process between agents, several patterns of hierarchies recur: (1) elite dominance, in which one or two leaders have a low esteem about all the other agents, who have negative esteems about each other except about the leaders; (2) a crisis pattern, in which every agent has a negative esteem about all others, including itself; (3) disorder, in which the agents have different perceived hierarchies; (4) an elite hierarchy for which all agents share a similar esteem about every other agent (called reputation), the reputations are widely spread between the hypothetical minimum and maximum (-1 and +1), and there are more agents of low reputation than of high reputation. The elite hierarchy model appears to be a classical stratification pattern with a wide base of low-prestige agents, with the number of agents in each strata of prestige progressively shrinking with higher ranks. Moreover, these four patterns exhibit different average signs of esteems, depending on whether agents gossip or not. In absence of gossip, esteems, except in the crisis pattern, can be of both signs (i.e., positive or negative). The hierarchy pattern is often composed from either positive and negative esteems, whereas the elite dominance pattern shows almost only negative esteems.

It should be noted, even if it is not the purpose of the present paper, that our study shows that with two types of agents having a different open-mindedness in the population, the only patterns of hierarchies which emerge are Elite dominance or Elite hierarchy. These patterns are characterized by high similarity between each agent's self-esteem level and its reputation.

\subsubsection{How gender inequality emerges from the agent-based model in our study}

The workings of previous studies using a similar model can help us understand how this differentiation emerges from the micro dynamics of our Leviathan model. The prior model \cite{Huet2017B, Huet2017} was composed by a population of agents with the same degree of open-mindedness. We showed the influence function, coupled with the noise during communication, causes agents' discussions increasing their self-esteems progressively, especially when their open-mindedness is large. Moreover the influence dynamics showed that agents not only became more positive about themselves, but also more negative towards others, due to their sensitivity to fluctuations of opinions in the population \cite{deffuant2018dark}. Indeed, while every agent in our study agrees, on average, on someone’s esteem, the noisy communications between agents make their esteems slightly different from this average. The maximum intensity of this difference is informed by the value of the noise parameter $\delta$ which is always small (i.e. 0.2 in our study). The sensitivity of an agent’s self-esteem and esteem for others to the noise implying fluctuations of esteem, depends on the agent’s level of open-mindedness. Indeed the lower this parameter is, the higher is the \guillemotleft positive ego-centric bias \guillemotright and the more negative is prejudice against other agents.

In the population studied in the present paper, we have a type of agents, men-like agents which listen to others less since they are more positively biased for ego and negatively biased for others because of their low open-mindedness. Thus, they perceive other agents more different from themselves than women-like agents, less biased, do. 

\begin{figure}
\includegraphics[scale=1]{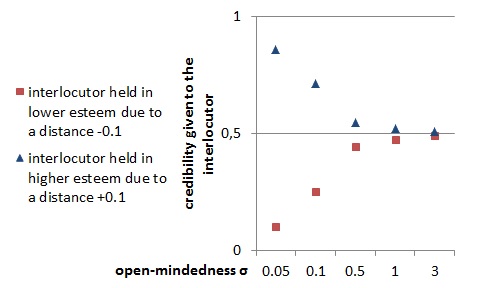}
 
\caption{\small{Credibility given to an interlocutor for whom our esteem varies around our self-esteem by 0.1 for different values of the open-mindedness $\sigma$; for a variation of: -0.1 implying the interlocutor is held in lower esteem (red square), or +0.1 implying the interlocutor is held is higher esteem (blue triangle).}}
\label{explanation}
\end{figure}

On the other hand, our two types of agents have a different open-mindedness. This difference leads agents, for a same fluctuation of esteem (0.1 in the example of the Fig. \ref{explanation}) to give a very different credibility to their interlocutor. Indeed, as shown in figure \ref{explanation}), one can observe than an agent with an open-mindedness of 0.05 will give a credibility of 0.86  to an agent perceived higher than her/himself by 0.1, and a credibility of 0.1 to an agent perceived lower than her/himself by 0.1. In contrast, an agent with an open-mindedness of 3 will almost give the same credibility 0.5 to the interlocutor varying in esteem from -0.1 to 0.1. This means that the noise received by the second agent will be considered in the same manner, whoever the source. On the contrary, the small open-mindedness agent is highly sensitive to the noise coming from someone held in higher esteem, but almost neglects the noise when it comes from a lower source. Practically, it means that low open-mindedness agents who reach high social ranks hardly must be affected by noisy communications. They are only sensitive to noise coming from people having higher social ranks. Their self-esteem then becomes highly stable as they obtain higher ranks because the number of agents they see as more credible than they is quite small at high ranks. On the contrary, women-like agents who reach a high self-esteem by chance are less susceptible to keep their self-esteem high since they are equally sensitive to the variations of what others say about them, whoever they are, lower or higher source. Their self-esteems are thus lower than the one of men-like agents, and less stable. 

To sum-up, this difference of stability for agents reaching intermediary rank of the social hierarchy gives men-like agents more chance to benefit from interactions with other agents, especially women-like agents which are less negatively biased toward them, to gain higher self-esteem. This same difference disfavors women-like agents, who ultimately have stronger chances to lose high social rank and the related high self-esteem. This is why we observe the differences of social ranks and chances to reach a higher rank when the current rank is high enough (see part of the results dedicated to the glass ceiling effect). This is also why we have globally observed, comparing our pure-gender populations to the mixed-gender populations, that men-like agents benefit from the interaction with women-like agents while women-like agents lose from it.

\bibliography{sample}
 
\section*{Acknowledgements}
 
We thank Ilaria Bertazzi for her advice and what she brings into this study about the glass ceiling effect.
  
\section*{Author contributions statement}

S.H., F.G. and F.P. wrote the paper and conceived of the model in relation to social sciences research about gender inequality. S.H. conceived the model and ran the simulation. S.H. and F.G. analyzed the data. F.G. performed the data visualization. 
 
%\section*{Additional information}
 
%To include, in this order: \textbf{Accession codes} (where applicable); \textbf{Competing interests} (mandatory statement).
 
%The corresponding author is responsible for submitting a \href{http://www.nature.com/srep/policies/index.html#competing}{competing interests statement} on behalf of all authors of the paper. This statement must be included in the submitted article file.
 
%\begin{figure}[ht]
%\centering
%\includegraphics[width=\linewidth]{stream}
%\caption{Legend (350 words max). Example legend text.}
%\label{fig:stream}
%\end{figure}
 
%\begin{table}[ht]
%\centering
%\begin{tabular}{|l|l|l|}
%\hline
%Condition & n & p \\
%\hline
%A & 5 & 0.1 \\
%\hline
%B & 10 & 0.01 \\
%\hline
%\end{tabular}
%\caption{\label{tab:example}Legend (350 words max). Example legend text.}
%\end{table}
 
%Figures and tables can be referenced in LaTeX using the ref command, e.g. Figure \ref{fig:stream} and Table \ref{tab:example}.
 
\end{document}